# Data Poisoning Attacks on EEG Signal-based Risk Assessment Systems

Zhibo Zhang
C2PS, Department of Electrical Engineering and Computer Science
Khalifa University
Abu Dhabi, United Arab Emirates
100060990@ku.ac.ae

Sani Umar
Department of Electrical Engineering and Computer Science
Khalifa University
Abu Dhabi, United Arab Emirates
100059856@ku.ac.ae

Ahmed Y. Al Hammadi
C2PS, Department of Electrical Engineering and Computer Science
Khalifa University
Abu Dhabi, United Arab Emirates
ahmed.yalhammadi@ku.ac.ae

Sangyoung Yoon
C2PS, Department of Electrical Engineering and Computer Science
Khalifa University
Abu Dhabi, United Arab Emirates
sangyoung.yoon@ku.ac.ae

Ernesto Damiani
C2PS, Department of Electrical Engineering and Computer Science
Khalifa University
Abu Dhabi, United Arab Emirates
ernesto.damiani@ku.ac.ae

Chan Yeob Yeun
C2PS, Department of Electrical Engineering and Computer Science
Khalifa University
Abu Dhabi, United Arab Emirates
chan.yeun@ku.ac.ae

*Abstract*—Industrial insider risk assessment using electroencephalogram (EEG) signals has consistently attracted a lot of research attention. However, EEG signal-based risk assessment systems, which could evaluate the emotional states of humans, have shown several vulnerabilities to data poison attacks. In this paper, from the attackers' perspective, data poison attacks involving label-flipping occurring in the training stages of different machine learning models intrude on the EEG signal-based risk assessment systems using these machine learning models. This paper aims to propose two categories of label-flipping methods to attack different machine learning classifiers including Adaptive Boosting (AdaBoost), Multilayer Perceptron (MLP), Random Forest, and K-Nearest Neighbors (KNN) dedicated to the classification of 4 different human emotions using EEG signals. This aims to degrade the performance of the aforementioned machine learning models concerning the classification task. The experimental results show that the proposed data poison attacks are model-agnostically effective whereas different models have different resilience to the data poison attacks.

## I. INTRODUCTION

Industrial insider risk [1] is a threat to an industrial organization, it originates from individuals who work there, such as current or former employees, and who have inside knowledge of the organization's security procedures, customer information, and computer systems. Therefore, it is significant to evaluate human behaviors with the help of artificial intelligence (AI) technologies [2]. Conventionally, facial [3] and speech [4] signals were utilized to analyze humans' emotions and behaviors. However, with some sort of training, these signals can be easily masked by humans [5]. Therefore, in recent years, electroencephalograms (EEGs) [6] have been utilized to evaluate the emotional state of humans to prevent industrial insider attacks as it is impossible for humans to camouflage or control their brainwaves. To improve the accuracy of identifying EEG signals, various risk assessment systems have used Machine Learning (ML) [7] classifiers in various configurations to analyze human emotions. On the other hand, the vulnerabilities of ML models [8] have been discovered and taken advantage of by the attackers to degrade the performance of the EEG signal-based risk assessment systems. To this end, attackers exploit data poisoning (DP) [9] attacks strategies to undermine the reliability of a target ML model by poisoning the ML models during the training stage.

Accordingly, this paper aims to craft DP attacks based on label-flipping, targeting the ML classifiers of the EEG signal-based risk assessment systems. Besides, different poisoning thresholds have been proposed in this paper to quantify the poisoning effects and vulnerabilities of each ML model [10]. In the future, Explainable Artificial Intelligence(XAI) [11] would be employed to examine and explain the exact influence of the DP attacks on the EEG signal-based risk assessment systems in terms of features and inner mechanisms [12].

In [5], the authors gathered data from 17 individuals in various emotional states, the collected data was charted and divided into four risk categories: low, normal, medium, and high. The Emotiv Insight EEG system, which has five electrodes, is used because it is intended to be economical. This paper intends to categorize unusual EEG signals pointing to a possible insider threat and whether the employee is qualified for duty.

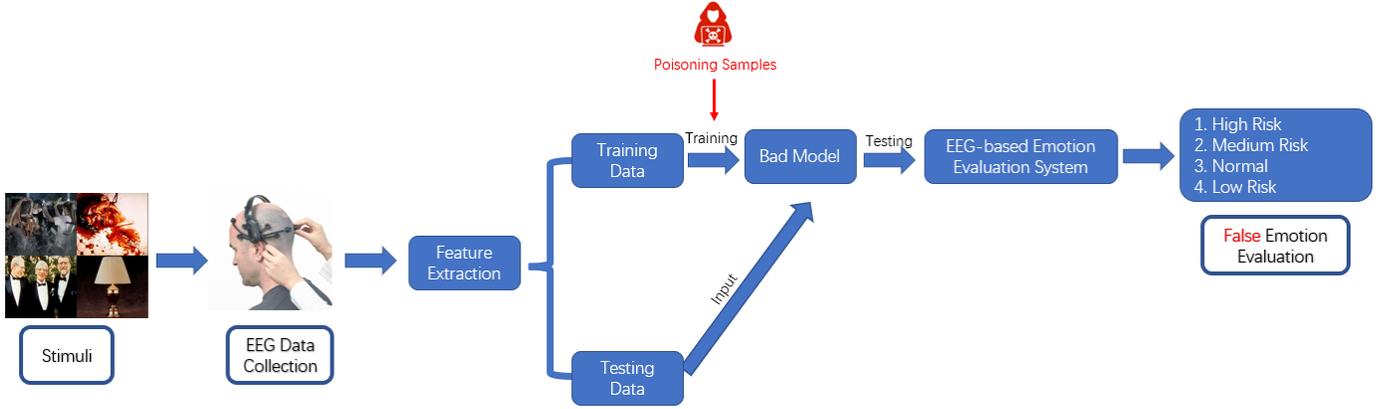

Fig. 1. The overview of the proposed data poison attacks on EEG-based risk assessment systems.

The most popular DEAP dataset [13], which contains EEG, is used by signals from 32 individuals, including 16 women and 16 males of various ages from 19 to 37 years old. The experiment to collect data was based on 40 musical videos played for the participants as stimuli. Every video was developed to elicit a particularly emotional experience. The dataset consists of total raw data of 1,280.

On the other hand, DP attacks targeting the ML models utilized by the EEG analysis systems were also proposed by researchers [14]. To the best of our knowledge, this paper makes the first-ever and practically feasible suggestion to employ a narrow period pulse to protect EEG-based brain-computer interfaces (BCIs) from poisoning attacks.

Therefore, to bridge the gap of deploying label-flipping DP attacks to ML models of EEG signal-based risk assessment systems from attackers' perspective, the main contributions of this paper are listed as follows:

(1) Applying different ML models in EEG signal-based risk assessment systems and evaluating human emotions.

(2) Deploying two different categories of label-flipping attacks in the training process of ML models in emotion evaluation tasks using ML models.

(3) Comparing and investigating the resilience of different ML models against label-flipping attacks.

The rest of this paper is organized as follows: Section II introduces the proposed label-flipping DP attacks to the ML models of EEG signal-based risk assessment systems. Section III provides experimentation results and analysis in terms of different evaluations of conventional performance metrics. Section IV concludes this paper and provides prospects for future work.

## II. METHODOLOGY

In this section, the methods including the data processing of EEG signals, different ML models utilized to evaluate human emotions based on EEG signals, and DP attacks based on label-flipping are introduced to build the framework of DP attacks on EEG signal-based risk assessment systems from the view of attackers. An overview of the proposed architecture is shown in Fig. 1 and the different parts of the proposed attacking framework are described in the subsections respectively.

### A. EEG Signal Collection and Processing

The collection of the EEG signal dataset was done in a dedicated lab at Khalifa University [15] by Ahmed Alhammadi. The dataset was gathered to look into the potential applications of brainwave signals to industrial insider threat identification. The Emotiv Insight 5 channels device was used to collect the dataset. Data from 17 people who agreed to participate in the data collection are included in the dataset.

To facilitate data processing, Emotiv Insight creates an EDF file that is then translated into a CSV file. Each signal for a taken image was sorted into the four risk categories, including High-Risk, Medium-Risk, Low-Risk and Normal, found in the risk matrix, and each signal was then labeled appropriately. The 26 characteristics in the data files consist of 25 inputs and 1 output. Theta, Alpha, Low Beta, High Beta, and Gamma are the five brainwave bands that are recorded by each of the EEG device's five electrodes.

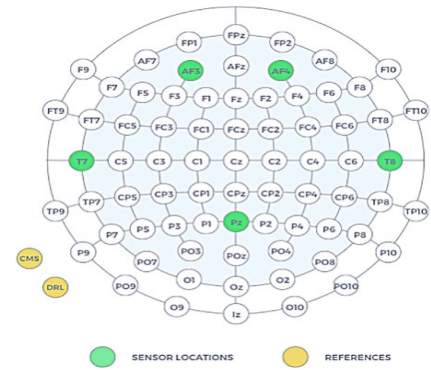

Fig. 2. Emotiv five channels positions.

TABLE I. LIST OF ALL INPUTS FROM EMOTIV ELECTRODES.

| Electrode | Input features |
|---|---|
| AF3 | AF3_THETA, AF_ALPHA, AF3_LOW_BETA, AF3_HIGH_BETA, AF3_GAMMA |

| Electrode | Input features |
|---|---|
| T7 | T7_THETA, T7_ALPHA, T7_LOW_BETA, T7_HIGH_BETA, T7_GAMMA |
| Pz | Pz_THETA, Pz_ALPHA, Pz_LOW_BETA, Pz_HIGH_BETA, Pz_GAMMA |
| T8 | T8_THETA, T8_ALPHA, T8_LOW_BETA, T8_HIGH_BETA, T8_GAMMA |
| AF4 | AF4_THETA, AF4_ALPHA, AF4_LOW_BETA, AF4_HIGH_BETA, AF4_GAMMA |

The electrodes in the Emotiv Insight device that record brainwaves associated with deceptive and cognitive activities are depicted in Fig. 2 (e.g., green dots). Making decisions based on emotional inputs, attributing others' intentions, and inferential reasoning are all related to AF3. Decision-making involving incentives and conflicts, planning, and judgment is under the purview of AF4. T7 and T8 deal with intentions, whereas Pz deals with cognitive processes. Therefore, the input data are shown in Table 1.

*B. Design of DP Attacks on ML Models of EEG Signal-based Risk Assessment Systems*

The DP attacks is performed on four different ML models namely Random Forest, AdaBoost, MLP, and KNN. These four ML classifiers were utilized for the classification tasks of risk assessment based on the collected EEG signal data without DP attacks initially. The dataset has been split into 80% for training the ML models, and 20% for testing.

As we discussed in the previous section, Label Flipping [16] technique has been proposed as the DP attack on the training sets of EEG signal data. Obviously, the most effective Label Flipping attacks are those that require the fewest poisonous samples; this has to do with the attacker's ability to inject poison samples in large quantities. The proportion of poisoning samples to the total number of samples in the training set figures as an important parameter to evaluate. And from the attackers' perspective, it is also valuable to evaluate the different resilience abilities of different ML models utilized in the classification based on the EEG signal data. Therefore, different poisoning rates which include 5% 25%, 50%, and 75% were performed on the training set and compared with the models' original performance (0% poisoning).

Furthermore, in this paper, two different Label Flipping scenarios are presented as there are four labels in the task of risk assessment system based on the EEG signal data. In the first scenario, other labels including Medium-Risk, Normal, and Low-Risk would be flipped into High-Risk. This is to observe the effect on the risk assessment system after falsifying other risks levels to a High-Risk level. On the other hand, in the second scenario, Low-Risk would be flipped into Normal, Normal would be flipped into Medium-Risk, Medium-Risk would be flipped into High-Risk, and High-Risk would be flipped into Low Risk. Similarly, the second scenario is applied in order to observe the effect on the EEG signal-based risk assessment system after falsifying risk levels to the other immediate risk levels.

### III. EXPERIMENT RESULTS AND ANALYSIS

As we discussed in the previous section, this paper utilized the "EEG Brainwave Dataset" [15] established by Ahmed Alhammadi at Khalifa University. There are 1550 samples in total, classified into High-Risk, Medium-Risk, Normal, and Low- Risk different risk categories. In this paper, 80% of samples in the dataset are utilized for training whereas 20% have been used for testing the four different ML models namely, AdaBoost, Random Forest, MLP, and KNN.

As for the DP attacks based on Label Flipping, attacks happen only in the training set. The poison rate includes 5%, 25%, 50%, and 75%. Moreover, we have used statistical metrics to evaluate the performance of the different ML models under the proposed DP attacks scenarios. The statistical metrics includes accuracy, recall, precision, and F1-score. Tables II and III presents the metrics comparison among different ML models under two different Label Flipping scenarios discussed in the previous section.

TABLE II. METRICS COMPARISON AMONG ML MODELS UNDER THE FIRST CATEGORY OF DP ATTACK.

| ML model | Poison rate [%] | Accuracy [%] | Recall [%] | Precision [%] | F1-Score [%] |
|---|---|---|---|---|---|
| Ada Boost | 0 | 99.68 | 99.67 | 99.67 | 99.66 |
| | 5 | 96.77 | 96.85 | 96.96 | 96.76 |
| | 25 | 73.87 | 74.67 | 61.68 | 65.99 |
| | 50 | 50.32 | 50.00 | 33.19 | 37.34 |
| | 75 | 24.19 | 25.00 | 6.05 | 9.74 |
| Random Forest | 0 | 90.97 | 91.03 | 91.94 | 91.04 |
| | 5 | 86.77 | 86.94 | 88.49 | 86.87 |
| | 25 | 60.97 | 61.93 | 55.66 | 55.33 |
| | 50 | 34.51 | 34.88 | 31.74 | 24.78 |
| | 75 | 24.19 | 25.00 | 6.05 | 9.74 |
| MLP | 0 | 76.13 | 75.97 | 76.24 | 75.67 |
| | 5 | 43.55 | 43.66 | 47.10 | 43.67 |
| | 25 | 42.26 | 43.38 | 39.99 | 35.99 |
| | 50 | 39.68 | 39.62 | 26.73 | 29.09 |
| | 75 | 24.19 | 25.00 | 6.05 | 9.74 |
| KNN | 0 | 80.65 | 80.57 | 80.48 | 80.43 |
| | 5 | 78.71 | 78.69 | 78.70 | 78.63 |
| | 25 | 57.42 | 58.11 | 48.35 | 51.31 |
| | 50 | 41.61 | 41.52 | 27.61 | 30.48 |
| | 75 | 24.19 | 25.00 | 6.05 | 9.74 |

TABLE III. METRICS COMPARISON AMONG ML MODELS UNDER THE SECOND CATEGORY OF DP ATTACK.

| ML model | Poison rate [%] | Accuracy [%] | Recall [%] | Precision [%] | F1-Score [%] |
|---|---|---|---|---|---|
| Ada Boost | 0 | 99.68 | 99.67 | 99.67 | 99.66 |
| | 5 | 96.77 | 96.78 | 96.79 | 96.74 |
| | 25 | 76.45 | 76.46 | 76.67 | 76.33 |
| | 50 | 57.74 | 58.25 | 63.59 | 56.61 |
| | 75 | 21.61 | 21.76 | 21.53 | 21.53 |
| Random Forest | 0 | 90.97 | 91.03 | 91.94 | 91.04 |
| | 5 | 88.71 | 88.85 | 89.87 | 88.79 |
| | 25 | 80.00 | 80.17 | 81.19 | 80.05 |
| | 50 | 50.00 | 50.56 | 57.41 | 48.02 |
| | 75 | 14.52 | 14.52 | 14.25 | 14.32 |
| MLP | 0 | 76.13 | 75.97 | 76.24 | 75.67 |
| | 5 | 43.55 | 43.92 | 46.16 | 43.42 |
| | 25 | 39.68 | 39.41 | 39.42 | 39.15 |
| | 50 | 34.84 | 35.41 | 39.80 | 30.73 |
| | 75 | 16.77 | 16.75 | 16.43 | 15.99 |
| KNN | 0 | 80.65 | 80.57 | 80.48 | 80.43 |
| | 5 | 79.35 | 79.31 | 79.39 | 79.18 |
| | 25 | 68.39 | 68.37 | 69.47 | 68.20 |
| | 50 | 50.00 | 50.38 | 49.92 | 48.56 |
| | 75 | 18.06 | 18.02 | 17.76 | 17.75 |

## IV. CONCLUSION AND FUTURE WORK

This paper introduced and assessed two categories of DP attacks based on Label Flipping on the ML models of the EEG Signal-based risk assessment systems. The ability of both attacks to significantly lower the overall accurate classification rate of various multi-class risk assessment classifiers has been proved. Whereas different ML classifiers show different resilences for different attacks. The optimal poison rates for attacking these ML models differ as well. In future work, XAI techniques will be employed to investigate the exact influence of DP attacks on the EEG signals and ML models of the risk assessment system to increase the robustness of EEG-based risk assessment systems against tampered training data, including a defense mechanism that can mitigate the damage of DP attacks.

## REFERENCES

[1] A. Y. Al Hammadi et al., "Novel EEG Sensor-Based Risk Framework for the Detection of Insider Threats in Safety Critical Industrial Infrastructure," IEEE Access, vol. 8, pp. 206222–206234, 2020, doi: 10.1109/ACCESS.2020.3037979.

[2] S.-K. Kim, C. Y. Yeun, E. Damiani, and N.-W. Lo, "A Machine Learning Framework for Biometric Authentication Using Electrocardiogram," IEEE Access, vol. 7, pp. 94858–94868, 2019, doi: 10.1109/ACCESS.2019.2927079.

[3] Y. Huang, J. Yang, P. Liao, and J. Pan, "Fusion of Facial Expressions and EEG for Multimodal Emotion Recognition," Comput. Intell. Neurosci., vol. 2017, p. e2107451, Sep. 2017, doi: 10.1155/2017/2107451.

[4] H.-K. Song et al., "Deep user identification model with multiple biometric data," BMC Bioinformatics, vol. 21, no. 1, p. 315, Jul. 2020, doi: 10.1186/s12859-020-03613-3.

[5] A. Y. Al Hammadi et al., "Explainable artificial intelligence to evaluate industrial internal security using EEG signals in IoT framework," Ad Hoc Netw., vol. 123, p. 102641, Dec. 2021, doi: 10.1016/j.adhoc.2021.102641.

[6] S. T. Aung et al., "Entropy-Based Emotion Recognition from Multichannel EEG Signals Using Artificial Neural Network," Comput. Intell. Neurosci., vol. 2022, p. e6000989, Oct. 2022, doi: 10.1155/2022/6000989.

[7] M.-P. Hosseini, A. Hosseini, and K. Ahi, "A Review on Machine Learning for EEG Signal Processing in Bioengineering," IEEE Rev. Biomed. Eng., vol. 14, pp. 204–218, 2021, doi: 10.1109/RBME.2020.2969915.

[8] N. Pitropakis, E. Panaousis, T. Giannetsos, E. Anastasiadis, and G. Loukas, "A taxonomy and survey of attacks against machine learning," Comput. Sci. Rev., vol. 34, p. 100199, Nov. 2019, doi: 10.1016/j.cosrev.2019.100199.

[9] A. Paudice, L. Muñoz-González, A. Gyorgy, and E. C. Lupu, "Detection of Adversarial Training Examples in Poisoning Attacks through Anomaly Detection." arXiv, Feb. 08, 2018. doi: 10.48550/arXiv.1802.03041.

[10] Z. Zhang et al., "Explainable Data Poison Attacks on Human Emotion Evaluation Systems based on EEG Signals." arXiv, Jan. 17, 2023. doi: 10.48550/arXiv.2301.06923.

[11] Z. Zhang, H. A. Hamadi, E. Damiani, C. Y. Yeun, and F. Taher, "Explainable Artificial Intelligence Applications in Cyber Security: State-of-the-Art in Research," IEEE Access, vol. 10, pp. 93104–93139, 2022, doi: 10.1109/ACCESS.2022.3204051.

[12] Z. Zhang, E. Damiani, H. A. Hamadi, C. Y. Yeun, and F. Taher, "Explainable Artificial Intelligence to Detect Image Spam Using Convolutional Neural Network," in 2022 International Conference on Cyber Resilience (ICCR), Oct. 2022, pp. 1–5. doi: 10.1109/ICCR56254.2022.9995839.

[13] S. Koelstra et al., "DEAP: A Database for Emotion Analysis ;Using Physiological Signals," IEEE Trans. Affect. Comput., vol. 3, no. 1, pp. 18–31, Jan. 2012, doi: 10.1109/T-AFFC.2011.15.

[14] L. Meng et al., "EEG-Based Brain-Computer Interfaces Are Vulnerable to Backdoor Attacks." arXiv, Jan. 02, 2021. doi: 10.48550/arXiv.2011.00101.

[15] A. A. AlHammadi, "EEG Brainwave Dataset." IEEE, Feb. 25, 2021. Accessed: Nov. 18, 2022. [Online]. Available: https://ieee-dataport.org/documents/eeg-brainwave-dataset

[16] Z. Hu, B. Tan, R. R. Salakhutdinov, T. M. Mitchell, and E. P. Xing, "Learning Data Manipulation for Augmentation and Weighting," in Advances in Neural Information Processing Systems, 2019, vol. 32. Accessed: Nov. 18, 2022. [Online]. Available: https://proceedings.neurips.cc/paper/2019/hash/671f0311e2754fcdd37f70a8550379bc-Abstract.html